\documentclass[aps,pra]{revtex4}

\usepackage{graphicx}
\usepackage{color}
\usepackage{amsmath}
\newcommand{\ket}[1]{\left| #1 \right\rangle}

\newcommand{\braket}[2]{\langle #1|#2 \rangle}

\begin{document}

\title{Quantum coherence and negative quasi probabilities\\ in a contextual three-path interferometer}

\author{Holger F. Hofmann}
\email{hofmann@hiroshima-u.ac.jp}
\affiliation{
Graduate School of Advanced Science and Engineering, Hiroshima University,
Kagamiyama 1-3-1, Higashi Hiroshima 739-8530, Japan
}

\begin{abstract}
Basic quantum effects are often illustrated using single particle interferences in two-path interferometers. A wider range of non-classical phenomena can be illustrated using three-path interferometers, but the increased complexity of quantum statistics in a three-dimensional Hilbert space makes it difficult to identify a representative set of observable properties that could be used to characterize specific phenomena. Here, I propose a characterization of pure states based on a five-stage interferometer recently introduced to demonstrate the relation between different measurement contexts (Optica Quantum 1, 63 (2023)). It is shown that the orthogonality relations between the states representing the different measurement contexts can be used to classify pure states within the three-dimensional Hilbert space according to the non-classical correlations between different contexts expressed by negative Kirkwood-Dirac distributions.
\end{abstract}

\maketitle

\section{Introduction}

Single particle interference phenomena provide some of the most accessible demonstrations of quantum effects, starting from fundamental investigations of the which-way problem \cite{Wot79,Scu91,Eng96} and the realization that even entanglement can be encoded in single photons \cite{Kwi97} and continuing with experimental realizations demonstrating fundamental aspects of quantum measurements \cite{Dur98,Wal02}. It was soon realized that interesting new physics can be observed as systems are scaled up, e.g. by using multiple paths as qudits \cite{Nev05}, or by applying complementarity and the quantum eraser to triple slits \cite{Sid15,Sha17}. In an important development, it was pointed out that three path interference can result in seemingly paradoxical effects when one traces the photon back through a nested interferometer \cite{Vai13,Dan13,Vai14}. These results inspired a fruitful discussion that has demonstrate the relevance of multi-path interference for fundamental issues in quantum mechanics \cite{Gri16,Eng17,Geb18,Yua19,Spo19,Han23}. These developments have been possible because all Hilbert spaces can be mapped onto a number of optical modes corresponding to the dimensionality of the Hilbert space, where each Hilbert space basis can be represented by the detection of a single particle in a corresponding set of output ports. The paradoxes originating from the which-way problem of two-path interference can then be related to the more fundamental problem of contextuality in quantum mechanics \cite{Wag24a}. The formal definition of quantum contextuality originated from a mathematical study of the relation between different measurements represented by non-commuting operators \cite{Spe60,Koc67} and later evolved into a number of practical scenarios highlighting the paradoxical relations between measurement outcomes predicted by the quantum formalism \cite{Can14,Li17,Qu21}. The work presented here is based on the relations introduced by Clifton \cite{Cli93} and later developed into a number of experimentally realizable scenarios \cite{Hua03,Lei05,Kly08,Bar09,Cab13}. Even though the physics of this scenario is quite simple, it is generally treated with the same level of abstraction that is characteristic of the field \cite{Pav25}. It is therefore useful to consider a theoretical approach that is closer to the actual physics of particle detection \cite{Ji24}. It is then possible to shift the focus of the discussion to the underlying physical mechanism that transforms one measurement context into another, which brings us back to the problem of multi-path interference. 

By combining the work on contextuality with the fundamental problem of which-path interference, I have recently introduced a three-path interferometer that illustrates quantum contextuality by demonstrating the impossibility of tracing the path of a photon through the interferometer for input states that violate a non-contextual inequality \cite{OpticaQ}. The paths in this interferometer can be represented by a set of ten states in the three-dimensional Hilbert space of a single photon passing through the interferometer, where the interference effects at the beam splitters are described by the inner products of the Hilbert space vectors of the paths going into the beam splitter with the paths that are transmitted or reflected. The contextual three-path interferometer thus provides a clear physical meaning to a large set of non-orthogonal states in a three-dimensional Hilbert space.
In this paper, I analyze the relation between the different paths in the interferometer based on the Hilbert space vectors that represent them. Since the coefficients of these vectors all have the same phases, it is possible to express all interference phenomena on the surface of a sphere, where the proximity of points indicates the possibility of joint control, while orthogonality expresses exclusivity. Of particular interest is the possibility of interferences between states of minimal nonzero overlap, since these interferences represent a counterintuitive relation between possibilities that very nearly exclude each other. A measure of the non-classicality for these relations can be obtained using quasi probabilities of the Kirkwood-Dirac type \cite{Kir33,Dirac}. Kirkwood-Dirtac distributions characterize the non-classical statistics experimentally observed in weak measurements and related techniques \cite{Lun11,Hof11,Lun12,Hof12,The17}. They also describe interference effects in optical systems \cite{Hal18}, characterizing both non-local and local coherences \cite{Iin18,Bud23}. It is therefore natural to consider Kirkwood-Dirac quasi probabilities as a prime candidate for the characterization of non-classical effects in discrete quantum systems and quantum circuits \cite{Los23,Ume24,Wag24b}, with a view towards the development of new quantum technologies \cite{Han24,Arv24}. In the following, I will focus on the Kirkwood-Dirac (KD) value as a joint statistical weight of two non-orthogonal states. The transition from positive to negative KD values happens at the lines that are orthogonal to one of the states related to each other by the KD value. It is therefore of interest to study the way in which these lines of orthogonality divide up the space of Hilbert space states on the surface of the sphere defined by the paths of the interferometer. The results show that, in addition to the ten path states, there are ten superposition states defined by their orthogonality to a pair of paths. Five of these states are already familiar from the formulation of quantum paradoxes \cite{Cli93,Cab13,Ji24}. The other five states seem to be consistent with sets of ``classical’’ particle paths that can be identified with their positive KD values. Starting from these states, it is possible to develop a complete characterization of the different interference phenomena that can be observed in the three-path interferometer. Although negative KD values are clearly concentrated at the pole where contextuality is directly observed in the form of an inequality violation \cite{Kly08,OpticaQ}, the only states without negative KD values on the Hilbert space sphere are the path eigenstates themselves. Whenever interference effects occur in each of the contexts, they are characterized by negative KD values that describe the unavoidable suppression of probabilities by destructive interference. The present analysis thus provides a comprehensive guide to the more complex interference effects that occur in three-dimensional Hilbert spaces and may therefore play an important role in advancing our intuitive understanding of quantum effects in larger systems.

The remainder of the paper is organized as follows. Section \ref{Sec:HS} introduces the three-dimensional Hilbert space of the three-path interferometer and the representation of its path states on a Hilbert space sphere. Section \ref{Sec:ortho} identifies the $\ket{N_i}$ states as states that violate a non-contextual inequality and contrasts them with the $\ket{\theta_k}$ states that do not violate any inequality and hence describe quasiclassical statistics. Section \ref{Sec:KD} introduces the Kirkwood-Dirac quasi probability and identifies the ten KD values that characterize quantum states in the three-path interferometer. Section \ref{Sec:class} introduces the classification of quantum states on the Hilbert space sphere based on the number and type of negative KD values. The surface of the Hilbert space sphere is divided into six classes with thirty-one sub-classes. In Section \ref{Sec:Nstate}, the KD values and the path statistics of the $\ket{N_i}$ states are related to each other. Since the  $\ket{N_i}$ states belong to three classes at once, their characteristics can provide insights into the relations between these classes and their sub-classes. In Section \ref{Sec:theta}, the same analysis is performed for the $\ket{\theta_k}$ states that sit at the intersection of four sub-classes. In Section \ref{Sec:extreme}, it is shown that the $\ket{\theta_k}$ states are close to the maximal negative KD values of the five KD values relating two outer paths of the interferometer to each other. Section \ref{Sec:joint} discusses states that achieve high path state fidelities in all five contexts at once. It is shown that an orthogonal basis can be constructed by selecting an appropriate combination of sub-classes on the quasiclassical belt of the Hilbert space sphere. Section \ref{Sec:conclude} summarizes the results and concludes the paper.

\section{The three path interferometer and its single photon Hilbert space}
\label{Sec:HS}

\begin{figure}[ht]
%%\vspace{-1cm}
\begin{picture}(500,220)
%%\put(0,0){\framebox(500,220){}}
\put(20,0){\makebox(400,220){\vspace{-3cm}
\scalebox{1}[1]{
\includegraphics{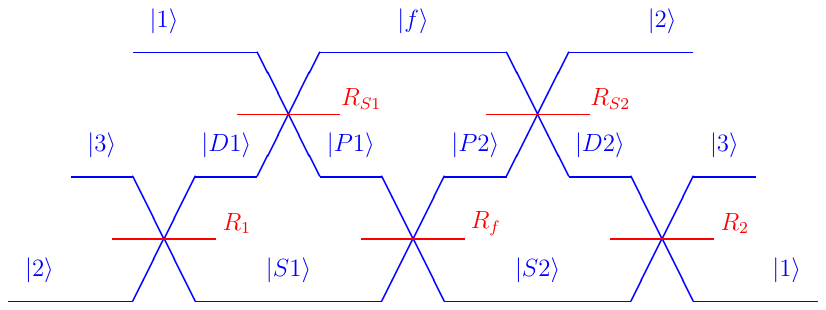}}}}
\end{picture}
%%\vspace{-5cm}
\caption{\label{fig1}
Contextual three-path interferometer. The reflectivities must be adjusted so that the output paths correspond to the input paths as shown. Here, I will consider the case of $R_1=R_2=1/2$, $R_{S1}=R_{S2}=1/3$, and $R_f=1/4$. 
}
\end{figure}

A three-path interferometer can be realized by pairwise interferences of the three paths, performed in sequence so that all paths interfere with each other. Incidentally, this allows a simple application of contextuality, since the unchanged third path is shared by the measurement contexts before and after the beam splitter. For the most simple example of contextuality in a three-dimensional Hilbert space, I have introduced a three-path interferometer with five different measurement contexts represented by ten orthogonal states \cite{Ji24,OpticaQ}. Fig. \ref{fig1} shows a schematic representation of the interferometer and the states corresponding to its path. The input basis is given by the states $\{\ket{1}, \ket{2}, \ket{3}\}$, where $\ket{3}$ is placed in the middle. In each subsequent interference, the middle path interferes with one of the two outer paths. Thus, the inner paths are short and belong to only one context each, while the outer paths are long and are shared by two different contexts each. The fifth context is identical to the input basis, establishing a cyclic symmetry of five possible measurements. It should be noted that the input basis is not privileged in any way, since contextuality is about the relation between different possible measurements. In the following discussion, the states $\ket{1}$ and $\ket{2}$ are included in the five outer paths, and the state $\ket{3}$ is included in the five inner paths. The resulting five-fold symmetry allows an efficient classification of interference phenomena across all five contexts.

In the following, I will consider an interferometer with reflectivities $R_1=1/2$, $R_{S1}=1/3$, $R_f=1/4$, $R_{S2}=1/3$, and $R_{2}=1/2$. Note that the index attached to each reflectivity indicates the path that runs parallel to the corresponding beam splitter. Consequently, the five outer paths can be expressed in the input basis as $\ket{1}$, $\ket{2}$, and
\begin{eqnarray}
\ket{S1} &=& \frac{1}{\sqrt{2}}(\ket{2} + \ket{3})
\nonumber \\
\ket{f} &=& \frac{1}{\sqrt{3}}(\ket{1}+\ket{2}-\ket{3})
\nonumber \\
\ket{S2} &=& \frac{1}{\sqrt{2}}(\ket{1} + \ket{3}),
\end{eqnarray}     
where the inner product of two outer path states is either zero or positive.
The five inner paths can be derived from orthogonality relations between the corresponding pair of outer paths. In sequence, they read $\ket{3}$ followed by
\begin{eqnarray}
\ket{D1} &=& \frac{1}{\sqrt{2}}(\ket{2} - \ket{3})
\nonumber \\
\ket{P1} &=& \frac{1}{\sqrt{6}}(2 \ket{1} - \ket{2} +\ket{3})
\nonumber \\
\ket{P2} &=& \frac{1}{\sqrt{6}}(-\ket{1} + 2 \ket{2}+\ket{3})
\nonumber \\
\ket{D2} &=& \frac{1}{\sqrt{2}}(\ket{1} - \ket{3}).
\end{eqnarray}   
Note that the inner product of two consecutive inner states is always negative, as clearly indicated by the negative sign of the component $\ket{3}$ in $\ket{D1}$ and $\ket{D2}$. Due to the cyclic symmetry of the contexts, we can use the expansion of states in the input basis to determine the sign and to estimate the relative magnitude of all other inner products. For instance, the inner product between $\ket{P1}$ and $\ket{P2}$ is similar to the inner product between $\ket{3}$ and $\ket{D1}$, while the inner product between $\ket{P1}$ and $\ket{D2}$ is similar to the inner product of $\ket{3}$ and $\ket{P1}$. In particular, the former are negative while the latter are positive. In addition, the former have higher values than the latter. To visualize these relations, it is best to consider the geometry of Hilbert space without any preferred basis. 

In total, we have ten different states in a three-dimensional Hilbert space, all related to each other by real-valued inner products. Each state can be represented by a three-dimensional vector of unit length, as shown by the expansions in the input basis $\{\ket{1}, \ket{2}, \ket{3}\}$. However, it should be noted that the sign of the total vector has no physical meaning, so each state appears twice, once each on opposite sides of the unit sphere. Two states share the same measurement context if they are orthogonal to each other. On the three-dimensional unit sphere, the states orthogonal to any given state lie on the great circle around the axis defined by the state. Every state can be defined as the intersection of two great circles corresponding to parallel paths in the interferometer. Since the five states corresponding to the inner paths of the interferometer belong to only one measurement context each, they are defined by their orthogonality to the two outer paths of that context. The five states that correspond to the outer paths belong to two different contexts and can be defined by their orthogonality to any combination of the two outer paths and the two inner paths of these two contexts. 

\begin{figure}[ht]
%%\vspace{-1cm}
\begin{picture}(500,400)
%%\put(0,0){\framebox(500,400){}}
\put(20,0){\makebox(400,400){%%\vspace{-0.5cm}
\scalebox{0.9}[0.9]{
\includegraphics{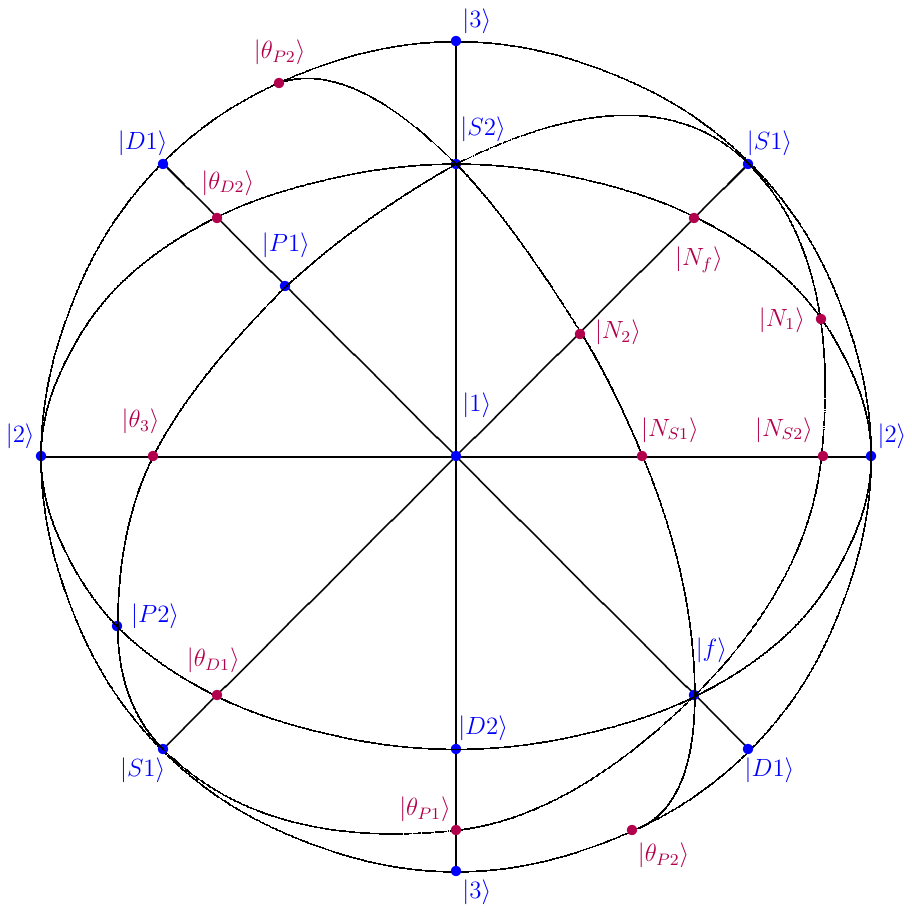}}}}
\end{picture}
%%\vspace{-5cm}
\caption{\label{fig2}
Illustration of Hilbert space directions corresponding to the ten paths in the contextual interferometer. Each great circle shown indicates orthogonality to one of the ten path states. Outer path states are found at the intersection of four great circles and inner path states are found at the intersection of two great circles. In addition, ten more states can be defined by orthogonality relations that are not satisfied by specific path states. 
}
\end{figure}

Fig. \ref{fig2} shows an illustration of the ten states and the great circles representing the orthogonality relations on the unit sphere of the three-dimensional real Hilbert space. Since each state appears twice on opposite sides of the sphere, it is sufficient to show only the hemisphere with positive values for the $\ket{1}$ component, using a projection into the plane spanned by $\ket{2}$ and $\ket{3}$ to represent the three-dimensional sphere.
As discussed above, the inner path states are orthogonal to two outer path states and are therefore located at intersections of the two great circles representing these states. The outer path states are orthogonal to two inner path states and two outer path states and are therefore located at intersections of four great circles. In addition, there are ten more intersections indicating states that are orthogonal to two paths from different contexts. These states are not path states and cannot be associated with any particular context. Instead, they are characterized by well-defined relations between different contexts that should be considered in more detail. 

\section{Orthogonality relations between different contexts}
\label{Sec:ortho}

The first set of non-path states is defined by orthogonality relations between the inner path states of two different contexts. These states are found at the corners of a pentagon formed by the great circles orthogonal to the five inner path states. This pentagon is part of a pentagram with the five outer path states at its points. Each of the states at the corners of the pentagon is opposite to one of the outer path states at the points of the pentagram. This geometry can be linked directly to a Hardy-like contextuality paradox \cite{Cli93,Cab13,Ji24,OpticaQ,Har93}, where quantum mechanics allows the population of a state with non-zero probabilties even though all paths that include the state run through states with a probability of zero. For instance, the state $\ket{N_f}$ is defined by $P(D1)=0$ and $P(D2)=0$. Since no photons are found in either $\ket{D1}$ or $\ket{D2}$, it would seem that photons traveling through $\ket{f}$ would have to pass from $\ket{1}$ to $\ket{2}$. However, these two states are orthogonal, so photons entering $\ket{f}$ from $\ket{1}$ should exit from $\ket{D2}$, and photons exiting from $\ket{2}$ should enter from $\ket{D1}$. $P(D1)=0$ and $P(D2)=0$ should therefore require that $P(f)$ be zero. Nevertheless quantum mechanics violates this expectation by requiring a non-zero probability of $P(f)=1/9$ for the path state $\ket{f}$ instead. Geometrically, this non-zero probability of $\ket{f}$ is indicated by the separation between the intersection of $P(D1)=0$ and $P(D2)=0$ on the one side, and the $P(f)=0$ line that runs between $\ket{S1}$ and $\ket{S2}$. Note that the same argument can be repeated for each of the five corners of the pentagon. Specifically, $\ket{N_1}$ has $P(1)=1/11$, $\ket{N_{S2}}$ has $P(S2)=1/10$, $\ket{N_{S1}}$ has $P(S1)=1/10$, and  $\ket{N_2}$ has $P(2)=1/11$. In terms of the $\{\ket{1}, \ket{2}, \ket{3}\}$ basis, the five states are given by 
\begin{eqnarray}
\ket{N_f} &=& \frac{1}{\sqrt{3}} \left(\ket{1}+\ket{2}+\ket{3}\right),
\nonumber \\
\ket{N_1} &=& \frac{1}{\sqrt{11}} \left(\ket{1}+ 3 \ket{2}+\ket{3}\right),
\nonumber \\
\ket{N_{S2}} &=& \frac{1}{\sqrt{5}} \left(\ket{1} + 2 \ket{2}\right),
\nonumber \\
\ket{N_{S1}} &=& \frac{1}{\sqrt{5}} \left(2 \ket{1}+ \ket{2}\right),
\nonumber \\
\ket{N_2} &=& \frac{1}{\sqrt{11}} \left(3 \ket{1}+ \ket{2}+\ket{3}\right).
\end{eqnarray}
Each of these five states violates the inequalities derived from the non-contextual assignments of photon paths \cite{Entropy}.

Non-contextual inequalities describe the statistical limits that an assignment of classical paths would impose on the detection probabilities in the different paths \cite{OpticaQ}. Since the detection probabilities add up to one in each of the five contexts, there are a number of equivalent ways to formulate this inequality. 
Although the most common formulation that is symmetric in all five contexts is the one inequality for the present paradox refers to the sum of the probabilities in the outer paths \cite{Kly08}, the contextual interferometer in Fig. \ref{fig1} suggests that the lower limit of inner path probabilities might be easier to understand. Each particle entering the interferometer in path $\ket{1}$ or $\ket{2}$ must pass through one of the inner paths to get to the corresponding output port. Therefore, the particle is either already in the inner path $\ket{3}$, or it must be found in one of the other four inner paths on its way from the input port to the output port. No matter what path a photon takes, it must always pass through at least one of the five inner paths. All non-contextual assignments of paths should therefore satisfy the inequality
\begin{equation}
\label{eq:ineq}
P(3)+P(D1)+P(P1)+P(P2)+P(D2) \geq 1.
\end{equation}
The five states given above all violate this inequality by a significant amount. For $\ket{N_f}$, the sum is $7/9$, leaving a probability of $2/9$ that the photon is never found in one of the inner paths. For $\ket{N_1}$ and $\ket{N_2}$, the missing probability is $2/11$, and for $\ket{N_{S1}}$ and $\ket{N_{S2}}$, it is $1/5$. 

The violation of non-contextual inequalities is a characteristic of all of the states inside the pentagon described by the five states $\ket{N_i}$. The maximal violation can be determined by diagonalizing the operator that describes the probability sum in Eq.(\ref{eq:ineq}). It is found at $P(1)=P(2)=0.4676$ and $P(3)=0.0648$, close to the middle of the pentagon, where the missing probability reaches its maximal value of $\sqrt{11/12}-1/2=0.4574$. The class of states enclosed by the five boundaries $P(3)=0$, $P(D1)=0$, $P(P1)=0$, $P(P2)=0$ and $P(D2)=0$ thus describes the quantum mechanical possibility of suppressing all five probabilities below the limit allowed by non-contextual theories. The proximity of the boundaries shows that the quantum statistics of the states are characterized by interference effects between low amplitudes corresponding to inner states and much larger amplitudes corresponding to outer states. This observation indicates that the interference effects involving the low amplitude components found in close proximity to the probability zero boundaries shown in Fig. \ref{fig1} modify quantum statistics in a highly non-trivial manner, even when no inequality violations are observed.

The general implications of the proximity of probability zero boundaries can be explored by considering the other set of states defined by zero probability of outcomes in two different contexts. These states, labeled $\ket{\theta_k}$ in Fig. \ref{fig2}, are orthogonal to the inner path $\ket{k}$ and the outer path from the contexts opposite to the context of this inner path. The statistics defined by these conditions depend strongly on the choice of the inner path $\ket{k}$, as can be seen from the distance between the states in Hilbert space. In terms of the $\{\ket{1}, \ket{2}, \ket{3}\}$ basis, the five states are given by 
\begin{eqnarray}
\ket{\theta_{P2}} &=& \frac{1}{\sqrt{5}} \left(\ket{2} - 2 \ket{3} \right),
\nonumber \\
\ket{\theta_{D1}} &=& \frac{1}{\sqrt{3}} \left(\ket{1} - \ket{2} - \ket{3}\right),
\nonumber \\
\ket{\theta_3} &=& \frac{1}{\sqrt{2}} \left(\ket{1}-\ket{2}\right),
\nonumber \\
\ket{\theta_{D2}} &=& \frac{1}{\sqrt{3}} \left(\ket{1} - \ket{2} + \ket{3}\right),
\nonumber \\
\ket{\theta_{P1}} &=& \frac{1}{\sqrt{5}} \left(\ket{1} - 2 \ket{3} \right).
\end{eqnarray}
The probability zero conditions of two neighboring $\theta$-states define the sides of a tetragon. For instance, the states $\ket{\theta_3}$ and $\ket{\theta_{D1}}$ are at opposite corners of the tetragon defined by $P(3)=0$, $P(f)=0$, $P(D1)=0$, $P(S2)=0$. The other two corners of this tetragon are given by $\ket{1}$ (orthogonal to $\ket{3}$ and $\ket{D1}$) and $\ket{P2}$ (orthogonal to $\ket{f}$ and $\ket{S2}$). It is worth noting that non-contextual assignments of photon paths can accommodate all four conditions at the same time. In fact,there are two different paths that are still possible even if the four paths defining the boundary are completely excluded – one given by the sequence $\{1, P1, P2, D2\}$ and the other by $\{2, S1,P2\}$. The states associated with this sequences are all located in the vicinity of the states $\ket{\theta_3}$ and $\ket{\theta_{D1}}$ in Hilbert space. 

In quantum mechanics, the proximity of states is an expression of uncertainty. Even if the probability of an outcome is high, full certainty is only possible when the state describing the measurement outcome is identical to the initial state. A complete characterization of the relation between different measurement contexts must therefore be based on statistical arguments. In the following, I will introduce the Kirkwood-Dirac quasi probability to identify the joint statistical weights of two non-orthogonal outcomes. These KD values can be negative, suggesting a relation that exceeds the mutual exclusion associated with orthogonality. Deviations from classical statistics are therefore associated with a suppression of probabilities that cannot be achieved with positive joint probabilities. As quasi probabilities, KD values characterize an arbitrary state $\ket{\psi}$ in terms of linear contributions to the detection probabilities in each of the paths. Negative KD values thus identify the characteristic statistics of a quantum state $\ket{\psi}$ and provide a useful indicator of the different kinds of quantum statistics that can be observed when the input state of the interferometer is given by a specific state located on the surface of the three dimensional Hilbert space sphere shown in Fig. \ref{fig2}.

\section{Joint statistics of quantum states}
\label{Sec:KD}

The Kirkwood-Dirac distribution was originally introduced as a general quasi probability representing the quantum analog of a joint probability distribution of two incompatible measurements \cite{Kir33,Dirac,Lun11,Hof11,Lun12,Hof12}. In the present scenario, the KD values will be used to describe non-classical correlations between pairs of measurement outcomes from different measurement contexts \cite{Entropy}. Although the Kirkwood-Dirac distribution is a complex-valued quasi probability, the present discussion concerns only states characterized by real valued Kirkwood-Dirac distributions. It is therefore convenient to define the KD values of a state $\ket{\psi}$ as
\begin{equation}
\varrho(a,b|\psi) = \mbox{Re}\left(\braket{b}{a} \braket{a}{\psi} \braket{\psi}{b} \right).
\end{equation}
The absolute value of the Kirkwood-Dirac term is directly related to the probabilities $P(a)$ and $P(b)$ by
\begin{equation}
|\varrho(a,b|\psi)| = \sqrt{P(a|\psi) P(b|\psi) |\braket{a}{b}|^2}.
\end{equation}
The value is zero for all states $\ket{\psi}$ on the line with $P(a)=0$ and along the line with $P(b)=0$. The absolute value increases as the state moves away from these lines. Since the inner product $\braket{a}{\psi}$ has opposite signs on the two sides of the $P(a)=0$ line, the areas enclosed by the $P(a)=0$ and $P(b)=0$ lines are characterized by the sign of their KD value $\varrho(a,b)$. On the Hilbert space sphere representing the quantum states of the three-path interferometer, these lines are great circles that divide up the sphere into four sectors, with the two narrower sectors having negative KD values and the two wider ones having positive KD values. The maximal absolute values are obtained at the center of each sector.  

As explained in \cite{Entropy}, the three-path interferometer can be characterized by a set of ten KD values, five of which combine an inner path with the outer path from the opposite context, and five of which combine two outer paths with each other. For brevity, I will refer to the former as inner KD values and to the latter as outer KD values. 
Specifically, the inner KD values are $\varrho(1,P2)$, $\varrho(S1,D2)$, $\varrho(f,3)$, $\varrho(S2,D1)$, $\varrho(2,P1)$, and the outer KD values are $\varrho(1,f)$, $\varrho(1,S2)$, $\varrho(S1,S2)$, $\varrho(2,S1)$, $\varrho(2,f)$. It may be worth noting that each of the KD values can be visualized by a path that connects both paths \cite{Entropy}. For outer KD values, there is only one inner path connecting the two outer paths. For inner KD values, the corresponding path includes the two additional short paths before and after the short path defined by the KD value.

The probabilities of each outer path can be expressed by a sum of two outer KD values and one inner KD value, corresponding to the three paths that connect the outer path to the three states of the complementary context, where all three states overlap with the initial outer path state. It should be noted that these relations hold for all possible states, since they correspond to the fundamental relation between a Kirkwood-Dirac quasi probability distribution and its marginal probabilities. Each of the KD value sums runs over a complete orthogonal basis of the three dimensional Hilbert space, where the specific basis represents a measurement context that is different from the one in which the probability of interest is observed. Putting the inner KD values last, the five sums are given by
\begin{eqnarray}
P(1|\psi) \;\; &=& \;\; \varrho(1,f|\psi) \;\;\, + \;  \varrho(1,S2|\psi) + \; \varrho(1,P2|\psi),
\nonumber \\
P(S1|\psi) &=& \varrho(S1,S2|\psi) + \; \varrho(2,S1|\psi) + \;\varrho(S1,D2|\psi),
\nonumber \\
P(f|\psi) \;\; &=& \;\; \varrho(1,f|\psi) \;\;\, + \;\; \varrho(2,f|\psi) \; + \;\; \varrho(f,3|\psi),
\nonumber \\
P(S2|\psi) &=& \varrho(S1,S2|\psi) + \; \varrho(1,S2|\psi) + \; \varrho(S2,D1|\psi),
\nonumber \\
P(2|\psi) \;\; &=& \;\; \varrho(2,f|\psi) \;\;\, + \; \varrho(2,S1|\psi) + \;\varrho(2,P1|\psi).
\end{eqnarray}
These relations show how the observable photon statistics can be obtained from the KD values of an arbitrary quantum state. They also confirm the intuitive notion that KD values represent joint probabilities associated with classical paths running through the interferometer. KD values thus provide a consistent description of the transmission of photons from one measurement context to another.

We can now identify the sectors of the Hilbert space sphere where a specific KD value is negative. The boundaries of these sectors are the great circles describing states that are orthogonal to one of the two paths of the KD value. For each inner KD value, these great circles intersect at the state $\ket{\theta_i}$, where $i$ is the inner path of the KD value. Inner KD values are negative in sectors that always include the pentagon formed by the five $\ket{N_i}$ states. For $\varrho(f,3)$, the boundaries of the negative sector run from $\ket{\theta_3}$ to $\ket{P1}$ to $\ket{S2}$ to $\ket{S1}$ to $\ket{P2}$ and back to $\ket{\theta_3}$, and from $\ket{\theta_3}$ to $\ket{1}$ to $\ket{N_{S1}}$ to $\ket{N_{S2}}$ to $\ket{2}$ and back to $\ket{\theta_3}$. As one would expect, the KD values of path states are always zero or positive. In the present case, path states orthogonal to $\ket{f}$ or $\ket{3}$ have a KD value of $\varrho(f,3)=0$ and are found at the boundaries of the negative sector. On the other hand, three of the five $\ket{N_i}$ states are located close to the center of the negative sector where the negative KD value achieves its maximal absolute value. For $\varrho(f,3)$, the three states with highly negative KD values are  $\ket{N_2}$, $\ket{N_f}$ and $\ket{N_1}$.  

For each outer KD value, the zero probability lines intersect at the outer path state that is orthogonal to both of the outer path states that define the KD value. The pentagon formed by the five $\ket{N_i}$ states is always located in the sectors with positive KD values. For $\varrho(S1,S2)$, the boundaries of the negative sector run from $\ket{f}$ to $\ket{D2}$ to $\ket{\theta_{D1}}$ to $\ket{P2}$ to $\ket{2}$ and back to $\ket{f}$, and from $\ket{f}$ to $\ket{1}$ to $\ket{P1}$ to $\ket{\theta_{D2}}$ to $\ket{D1}$ and back to $\ket{f}$. Close to the center of the sector is the $\ket{\theta_3}$ state, indicating that the KD value of this state is close to the maximal negative value for $\varrho(S1,S2)$. In general, each of the five $\ket{\theta_k}$ states is close to the maximal negative  value for one of the five outer KD values.

\section{Classification of states based on negative KD values}
\label{Sec:class}

The zero probability lines divide the Hilbert space sphere of the three-path system into classes of states that are characterized by a specific distribution of negative KD values. Each class is defined by a set of polygons that can be identified by the states at their corners. In this section, I will develop a classification of all states on the sphere of real-valued Hilbert space vectors by identifying the characteristics of each of the 31 polygons into which the zero probability lines divide the Hilbert space sphere shown in Fig. \ref{fig2}. Based on the number of negative inner KD values and negative outer KD values, it is possible to distinguish six different classes of states.

\begin{table}[ht]
    \centering \large
    \begin{tabular}{|cc||cc|cc||}
    \multicolumn{6}{c}{}
    \\ \hline
       Class &&\multicolumn{3}{c}{negative KD values}& \\ 
       \hspace{2cm} && \hspace{0.5cm} inner \hspace{0.5cm} &&
        \hspace{0.5cm} outer \hspace*{0.5cm}&
        \\ \hline \hline 
        &&&&&
        \\[-0.3cm]
        $N$ && 5 && 0 &
        \\
        $V$ && 4 && 0 &
        \\
        $B$ && 3 && 0 &
        \\
        $T$ && 2 && 2 &
        \\
        $X$ && 1 && 2 &
        \\
        $Q$ && 0 && 2 &
        \\ 
        &&&&&
        \\
        \hline \hline
    \end{tabular}
    \caption{\label{tab1} Classes of three-path states according to their negative KD values. The number of negative inner KD values is sufficient to identify each class, while the number of negative outer KD values is either zero or two.}
\end{table}

{\bf Class $N$} is the class of states in the pentagon defined by the five states $\{\ket{N_f}, \ket{N_1}, \ket{N_{S2}}, \ket{N_{S1}}, \ket{N_{2}}\}$. As mentioned above, all states in this class violate  non-contextual inequalities. For states in this class, all five inner KD values are negative, while all of the outer KD values are positive.  

{\bf Class $V$} is the class of states that are found in the triangles with two states $\ket{N_i}$ at the base and one of the outer path states at the top. There are five such triangles, and the class can be divided into sub-classes $V(i)$, where $i=f,1,S2,S1,2$ denotes the outer path state at the corner of the triangle. For $V$ states, four of the inner KD values are negative. The positive KD value is the one associated with the states defining the top corner and the base, e.g. $\varrho(f,3)$ for $V(f)$ or $\varrho(2,P1)$ for $V(2)$. 

{\bf Class $B$} is the class of states in the triangles defined by one state $\ket{N_i}$ and two outer path states. There are five triangles of this kind, and the bases of the triangles form a pentagon with the five outer path states at its corners. The class can be divided into sub-classes $B(i,j)$, where $(i,j)=(1,f), (1,S2), (S1,S2), (2,S1), (2,f)$ are the pairs of outer path states at the base of each triangle. For these $B$ states, three of the inner KD values are negative. The positive KD values are those inner KD values that are associated with the two outer path states that define the sub-class, e.g. $\varrho(1,P2)$ and $\varrho(f,3)$ for $B(1,f)$ or $\varrho(S1,D2)$ and $\varrho(S2,D1)$ for $B(S1,S2)$.  

{\bf Class $T$} is the class of states in the triangles defined by two outer path states and one inner path state. Each of these combinations of states corresponds to a sequence of paths that could describe a classical particle trajectory running through the interferometer. Similarly to class $B$, this class can be divided into sub-classes $T(i,j)$, where $(i,j)=(1,f), (1,S2), (S1,S2), (2,S1), (2,f)$ are the pairs of outer path states at the base of each triangle. For the $T$ states, only two of the inner KD values are negative, but there are also two negative KD values among the outer KD values. The negative outer KD values are the ones associated with the outer path state orthogonal to the two outer path states, e.g. $\varrho(1,f)$ and $\varrho(2,f)$ for $T(S1,S2)$ or $\varrho(2,f)$ and $\varrho(2,S1)$ for $T(1,S2)$. The negative inner KD values associated with inner paths that are orthogonal to the outer path that defines the negative outer KD values. For $T(S1,S2)$, these are $\varrho(1,P2)$ and $\varrho(2,P1)$, and for $T(1,S2)$, they are $\varrho(S1,D2)$ and $\varrho(f,3)$. 

{\bf Class $X$} is the class of states in the triangles defined by an outer path state, an inner path state, and a state $\ket{\theta_k}$. This class can be divided into ten sub-classes $X(i,k)$, where $i$ is the outer path and $k$ is the inner path. Each sub-class is adjacent to one of the five sub-classes of $B$, where the outer path is shared with the sub-class $B(i,j)$, and the inner path is the one that is neither orthogonal to $i$ nor to $j$. At this boundary between $X$ and $B$, the inner KD value associated with the outer path that is orthogonal to $\ket{i}$ changes its sign, leaving only one negative inner KD value, e.g. $\varrho(1,P2)$ for $X(S2,3)$ or $\varrho(f,3)$ for $X(1,P1)$. The negative outer KD values also change at the boundary, flipping the signs of the KD values associated with the same outer state. For $X(S2,3)$, these are the outer KD values associated with $\ket{2}$, so $\varrho(1,f)$ remains negative and $\varrho(2,S1)$ is added. For $X(1,P1)$, these are $\varrho(2,f)$ and $\varrho(S1,S2)$.

{\bf Class $Q$} is the class of states in the quadrilaterals defined by two $\ket{\theta_k}$ states, one outer state and one inner state. This class can be divided into five sub-classes $Q(i,l)$, where $i$ is the outer path at the corner of the quadrilateral, and $l$ is the inner path at the corner of the quadrilateral. At the two boundaries defined by the outer path and one of the $\ket{\theta_i}$ states, the states are orthogonal to one of the inner path states, and this inner path state is associated with the negative inner KD value of the $X(i,k)$ state on the other side of the boundary. The only difference between the KD values of the $Q(i,l)$ class of states and the adjacent $X(i,k)$ class is that all of the inner KD values of $Q$-class states are positive. The two negative outer KD values are shared by the $Q(i,l)$ sub-class and the neighboring $X(i,k)$ and $X(i,k^\prime)$ sub-classes. For example, the negative values of $\varrho(1,f)$ and $\varrho(2,S1)$ are shared by $X(S2,3)$, $Q(S2,D1)$, and $X(S2,P1)$. Likewise, the negative values of $\varrho(2,f)$ and $\varrho(S1,S2)$  are shared by $X(1,P1)$, by $Q(1,P2)$, and by $X(1,D2)$.

Table \ref{tab1} summarizes the classes and their negative KD values. It is worth noting that the minimal number of negative KD values for all of the classes is two. Negative KD values are a natural consequence of the Hilbert space description of quantum statistics. The transition between classes happens at boundaries where several KD values are zero. These KD values can be counted as negative or as positive, allowing the states to belong to several classes at once. In the case of the path states, this means that all of the negative KD values are actually zero. The outer path states exist at intersections of the five classes $V$, $B$, $T$, $X$ and $Q$, satisfying their criteria with four inner KD values of zero and three outer KD values of zero. For the inner path states, two of the inner KD values and four of the outer KD values are zero, satisfying the criteria of classes $T$, $X$ and $Q$. 

The location of path states at an intersection of multiple classes results in positive KD values that correspond to a classical distribution of trajectories through the interferometer. The situation is different for the $\ket{\theta_k}$ and $\ket{N_i}$ states. The $\ket{\theta_k}$ states exist at intersections of the two classes $Q$ and $X$. One outer KD value is negative in all four sub-classes that the state belongs to, so that the $\ket{\theta_k}$ has a single negative KD value, as mentioned in sec. \ref{Sec:KD} above. The transitions between the sub-classes are described by values of zero for two outer KD values and one inner KD value. The $\ket{N_i}$ states exist at intersections of three classes $N$, $V$ and $B$. This requires only two inner KD values of zero, while the remaining three inner KD values must all be negative. Even though the number of negative KD values is rather large, the zero probability conditions make it comparatively easy to analyze the statistics of the $\ket{N_i}$ states. In the next section, I will discuss the relation between the KD values and the path probabilities of the $\ket{N_i}$ states in detail.

\section{Statistics of the $\ket{N_i}$ states}
\label{Sec:Nstate}

The states $\ket{N_i}$ exist at the boundaries of class N with two sub-classes of class V and one sub-class of class B. Three of their inner KD values are negative, and the other two are zero. Each state $\ket{N_i}$ is nearly orthogonal to the corresponding outer path state $\ket{i}$, and the two outer KD values of $\ket{i}$ are expected to have comparatively low values. On the other hand, each state $\ket{N_i}$ is closest to two outer path states, and these two paths achieve the highest probabilities within the interferometer. 

Even though some of the KD terms are negative, it is possible to identify a very simple relation between the observable probabilities of the paths and the KD values for each $\ket{N_i}$ state. For the $\ket{N_2}$ state, the following three KD values can be related directly to path probabilities, 
\begin{eqnarray}
\label{eq:corresp}
\varrho(1,S2|N_2) &=& P(P1|N_2),
\nonumber \\
\varrho(S1,S2|N_2) &=& P(S1|N_2),
\nonumber \\
\varrho(1,f|N_2) &=& P(f|N_2).
\end{eqnarray}
These three KD values all relate to the two outer paths $\ket{1}$ and $\ket{S2}$ that are orthogonal to $\ket{2}$. Each KD value is given by the probability of the paths in the context $\{\ket{f},\ket{S1},\ket{P1}\}$ that is part of the sequence of paths defined by the two arguments of the KD value. In the case of $\varrho(1,S2)$, this is the path $P1$ that connects $1$ and $S2$ through the beam splitters marked $R_{S1}$ and $R_f$ in Fig. \ref{fig1}. Note that similar relations exist for all $\ket{N_i}$ states. 

Eq.(\ref{eq:corresp}) suggests a classical distribution over three possible path sequences in the interferometer. However, this distribution cannot explain the non-zero probability $P(2|N_1)$ of finding the photon in path $\ket{2}$. This probability can be related to the three $KD$ values associated with $\ket{2}$. Specifically,
\begin{eqnarray}
%%\label{eq:corresp}
\varrho(2,S1|N_2) &=& P(2|N_2),
\nonumber \\
\varrho(2,f|N_2) &=& P(2|N_2),
\nonumber \\
\varrho(2,P1|N_2) &=& -P(2|N_2).
\end{eqnarray}
Here, the negative inner KD value $\varrho(2,P1)$ compensates one of the two positive outer KD values, so that an absolute value equal to the sum of the three KD values can be attributed to each. However, the two outer values also contribute to the probabilities $P(S1)$ and $P(f)$, respectively. To compensate these contributions, the two remaining inner KD values must have the same negative value as $\varrho(1,P2)$,
\begin{eqnarray}
%%\label{eq:corresp}
\varrho(S1,D2|N_2) &=& -P(2|N_2),
\nonumber \\
\varrho(f,3|N_2) &=& -P(2|N_2).
\end{eqnarray}
It is thus possible to separate the statistics given by the KD values into a positive quasiclassical set of values given in Eq. (\ref{eq:corresp}) and a set of positive and negative KD values related to the paradoxical appearance of a probability of $P(2|N_2)=1/11$. 

\begin{table}[ht]
    \centering \large
    \begin{tabular}{|c||c|c|c|c|c|}
    \multicolumn{6}{c}{}
    \\ \hline \hspace{2cm} &&&&& \\[-0.3cm]
       State &$\;\varrho(1,P2)\;\;$&$\varrho(S1,D2)$
       &$\;\;\varrho(f,3)\;\;$&$\varrho(S2,D1)$&$\;\varrho(2,P1)\;$ 
        \\[0.2cm] \hline \hline 
        &&&&&
        \\[-0.3cm]
        $\ket{N_1}$ & -1/11 & 0 & -1/11 & -1/11 & 0
        \\
        $\ket{N_{S1}}$ & 0 & -1/10 & 0 & -1/10 & -1/10
        \\
        $\ket{N_f}$ & -1/9 & 0 & -1/9 & 0 & -1/9
        \\
        $\ket{N_{S2}}$ & -1/10 & -1/10 & 0 & -1/10 & 0
        \\
        $\ket{N_2}$ & 0 & -1/11 & -1/11 & 0 & -1/11
        \\ 
        &&&&&
        \\
        \hline \hline
    \end{tabular}
    \caption{\label{tab2} Negative KD values of the five $\ket{N_i}$ states. Each state has two inner KD values of zero, indicating the $V$ and $B$ sub-classes that the state belongs to in addition to class $N$.}
\end{table}

Regarding the classes of states introduced in this paper, the $\ket{N_2}$ state is part of the N-class, where all inner KD values are zero or negative. This class is symmetric in the five contexts, with each $\ket{N_i}$ state representing a maximal bias within the class. The inner KD values of the five $\ket{N_i}$ states are shown in Table \ref{tab2}.
The bias can be characterized by the other classes that the state is part of. The $\ket{N_2}$ state is part of the $V(1)$ and $V(S2)$ sub-classes of the $V$ class. The $V$-class is characterized by a single positive inner KD term, associated with the state that defines the sub-class. In the $V(1)$ subclass, $\varrho(1,S2)$ and $\varrho(1,f)$ are high, and $\varrho(1,P2)$ is positive. Likewise, the $V(S2)$ subclass has high values for $\varrho(1,S2)$ and $\varrho(S1,S2)$, with a positive value of $\varrho(S2,D1)$. Finally, $\ket{N_2}$ is also part of the $B(1,S2)$ sub-class of the $B$ class of states. This sub-class has positive values for $\varrho(1,P2)$ and $\varrho(S2,D1)$, and is characterized by a high value of $\varrho(1,S2)$, with a maximum at the $P(2)=0$ boundary to the $T(1,S2)$ class. 

In summary, each of the five $\ket{N_i}$ states describes a quasiclassical distribution over three KD values determined by the two outer path states orthogonal to $\ket{i}$ and hence closest to $\ket{N_i}$. The quasiclassical statistics of $\ket{N_2}$ are given by
\begin{eqnarray}
\varrho(1,S2|N_2) &=& 6/11,
\nonumber \\
\varrho(S1,S2|N_2) &=& 2/11,
\nonumber \\
\varrho(1,f|N_2) &=& 3/11.
\end{eqnarray}
For comparison, the corresponding values for the neighboring state $\ket{N_{S1}}$ read
\begin{eqnarray}
\varrho(1,f|N_{S1}) &=& 2/5 = P(D2|N_{S1}),
\nonumber \\
\varrho(2,f|N_{S1}) &=& 1/5 = P(2|N_{S1}), 
\nonumber \\
\varrho(1,S2|N_{S1}) &=& 2/5 = P(S2|N_{S1}), 
\end{eqnarray}
and on the other side, there is $\ket{N_{f}}$ with
\begin{eqnarray}
\varrho(S1,S2|N_f) &=& 1/3 = P(3|N_f),
\nonumber \\
\varrho(2,S1|N_f) &=& 1/3 = P(2|N_f), 
\nonumber \\
\varrho(1,S2|N_f) &=& 1/3 = P(1|N_f). 
\end{eqnarray}
Each $\ket{N_i}$ state is defined by its proximity to two outer paths. These two outer states are included in the sub-class of the $B$ class that each $\ket{N_i}$ state belongs to. They are also part of the $P(i)=0$ boundary. It may be worth noting that non-contextual logic would require that the intersection between the zero probability lines that defines $\ket{N_i}$ should always coincide with the $P(i)=0$ boundary. The $B$ class thus represents the difference between Hilbert space relations and classical non-contextual reasoning. 

It is possible to identify coherent states corresponding to maximal probabilities along a trajectory through two adjacent outer paths by considering the states along the $P(i)=0$ boundary connecting the two outer path states. This boundary separates $B$ class states from $T$ class states. The maximal KD value for two adjacent outer paths is thus found at the center of a quadrilateral formed by combining the corresponding sub-classes of $B$ and $T$. For instance, $B(1,S2)$ and $T(1,S2)$ form a quadrilateral with $\ket{N_2}$, $\ket{1}$, $\ket{S2}$ and $\ket{P1}$ at the corners. These states are characterized by high KD values for $\varrho(1,S2)$. They are also characterized by low probabilities of $\ket{2}$, $\ket{S1}$, $\ket{f}$, $\ket{D1}$ and $\ket{P2}$. For the state $\ket{N_2}$, the suppression of $\ket{S1}$ and $\ket{f}$ is minimal, as expressed by the corresponding KD values of $\varrho(S1,S2)=P(S1)$ and $\varrho(1,f)=P(f)$. The contrast between the different contexts is lowest at the $\ket{N_i}$ states and increases towards the $T$ class states. 

\section{Statistics of the $\ket{\theta_k}$ states}
\label{Sec:theta}

We can now apply the same method of analysis to the $\ket{\theta_k}$ states. These states exist at the boundaries between $Q$ class and $X$ class states, where each $\ket{\theta_k}$ state is part of two $Q$ sub-classes and two $X$ sub-classes. One of the outer KD values is negative and two are zero. Of the inner KD values, four are positive and only one is zero. It should be noted that two neighboring $\ket{\theta_k}$ states are part of the same $Q(i,k)$ subclass characterized by the two negative outer KD values of these two states. 

The $\ket{\theta_k}$ states do not violate any inequality, indicating that their statistics can be explained by a hidden variable model assigning positive probabilities to specific combinations of paths. Such hidden variable models correspond to a set of positive KD values with a total probability of one. For the $\ket{\theta_k}$ states, this is a set of four KD values that are directly related to corresponding path probabilities. For the $\ket{\theta_{D1}}$ state, the four KD values are
\begin{eqnarray}
\label{eq:thetacorr}
\varrho(2,S1|\theta_{D1}) &=& P(2|\theta_{D1}),
\nonumber \\
\varrho(1,f|\theta_{D1}) &=& P(f|\theta_{D1}),
\nonumber \\
\varrho(S1,D2|\theta_{D1}) &=& P(3|\theta_{D1}),
\nonumber \\
\varrho(1,P2|\theta_{D1}) &=& P(P1|\theta_{D1}).
\end{eqnarray}
The sum of these four probabilities is one, leaving no part of the statistics unexplained. The contributions of the remaining KD values that would appear in the outer path probabilities must all cancel out. For the negative KD value of $\ket{\theta_{D1}}$,
\begin{eqnarray}
\varrho(2,f|\theta_{D1}) &=& - \varrho(2,P1|\theta_{D1})
\nonumber \\
 &=& - \varrho(f,3|\theta_{D1}).
\end{eqnarray} 
It may be worth noting that these three KD values would have to be zero in any hidden variable theory, since they correspond to photon trajectories that include $\ket{D1}$ even though $P(D1)=0$. For example, photons passing through $\ket{f}$ and $\ket{3}$ must pass through both $\ket{D1}$ and $\ket{D2}$, and photons passing through $\ket{2}$ and $\ket{P1}$ must pass through both $\ket{D1}$ and $\ket{P2}$. To find the magnitude of the negative KD value, one can use the relation 
\begin{equation}
(\varrho(i,j))^2 = | \braket{i|j} |^2 P(i) P(j).
\end{equation}
For the $\ket{\theta_{D1}}$ and $\ket{\theta_{D2}}$ states, the result is $1/9$, for the $\ket{\theta_{P1}}$ and $\ket{\theta_{P1}}$ states, it is $1/10$, and for the $\ket{\theta_3}$ state, it is $1/8$. 

Each $\ket{\theta_k}$ state is part of four sub-classes of states. For the $\ket{\theta_{D1}}$ state, these are the $Q(S1,D2)$ and the $Q(1,P2)$ sub-classes of the $Q$ class and the $X(1,D2)$ and the $X(S1,P2)$ sub-classes of the $X$ class. The $Q(S1,D2)$ and $X(1,D2)$ sub-classes have negative values for $\varrho(2,f)$ and $\varrho(1,S2)$, while the $Q(1,P2)$ and $X(S1,P2)$ sub-classes have negative values for $\varrho(2,f)$ and $\varrho(S1,S2)$. Moving into $Q(S1,D2)$ increases $\varrho(S1,D2)$, moving into $X(1,D2)$ increases $\varrho(1,f)$, moving into $Q(1,P2)$ increases $\varrho(1,P2)$, and moving into $X(S1,P2)$ increases $\varrho(S1,S2)$. Each of the four dominant positive KD values can thus be identified with one of the four sub-classes that the state $\ket{\theta_{D1}}$ is part of.  

In summary, each of the five $\ket{\theta_k}$ states describes a quasiclassical distribution over four KD values determined by the corresponding set of path probabilities. For the state $\ket{\theta_{D1}}$, these KD values are
\begin{eqnarray}
\label{eq:thetaD1}
\varrho(2,S1|\theta_{D1}) &=& 1/3,
\nonumber \\
\varrho(1,f|\theta_{D1}) &=& 1/9,
\nonumber \\
\varrho(S1,D2|\theta_{D1}) &=& 1/3,
\nonumber \\
\varrho(1,P2|\theta_{D1}) &=& 2/9.
\end{eqnarray}
For comparison, the corresponding values for the neighbouring state $\ket{\theta_{3}}$ read
\begin{eqnarray}
\varrho(1,S2|\theta_3) &=& 1/4 = P(S2|\theta_3),
\nonumber \\
\varrho(2,S1|\theta_3) &=& 1/4 = P(S1|\theta_3),
\nonumber \\
\varrho(1,P2|\theta_3) &=& 1/4 = P(D2|\theta_3),
\nonumber \\
\varrho(2,P1|\theta_3) &=& 1/4 = P(D1|\theta_3).
\end{eqnarray}
Both $\ket{\theta_{D1}}$ and $\ket{\theta_{3}}$ are part of the $Q(1,P2)$ sub-class. They share high positive contributions to $\varrho(1,P2)$ and $\varrho(2,S1)$. On the other side of $\ket{\theta_{D1}}$ we find the state $\ket{\theta_{P1}}$ with
\begin{eqnarray}
\varrho(1,f|\theta_{P1}) &=& 1/5 = P(1|\theta_{P1}),
\nonumber \\
\varrho(S1,S2|\theta_{P1}) &=& 1/10 = P(S2|\theta_{P1}),
\nonumber \\
\varrho(f,3|\theta_{P1}) &=& 2/5 = P(D1|\theta_{P1}),
\nonumber \\
\varrho(S1,D2|\theta_{P1}) &=& 3/10 = P(P2|\theta_{P1}).
\end{eqnarray}
Both $\ket{\theta_{D1}}$ and $\ket{\theta_{P1}}$ are part of the $Q(S1,D2)$ sub-class. They share high positive contributions to $\varrho(S1,D2)$ and $\varrho(1,f)$. The four positive KD values of the state $\ket{\theta_{D1}}$ shown in Eq.(\ref{eq:thetaD1}) are all shared with the neighboring states $\ket{\theta_3}$ and $\ket{\theta_{P1}}$, supporting the quasiclassical interpretation of quantum states as overlapping probability distributions over the various classical paths. Maximal values of inner KD values can be obtained between two neighboring states $\ket{theta_k}$. These states are found in a quadrilateral formed by the inner and outer path that define the KD value and the two $\ket{\theta_k}$ states. The states inside this quadrilateral belong to a sub-class of $Q$ with only two negative KD values. In the following, I will take a closer look at the relation between negative and positive KD values based on the classification of states on the Hilbert space sphere.

\section{Extremal KD values}
\label{Sec:extreme}

Non-contextual inequalities are violated when the inner KD values are sufficiently negative \cite{OpticaQ}. Both the maximal inequality violation and the maximal negative inner KD values are found within the $N$ class of states. As the classification of states shows, negative outer KD values do not contribute to the inequality violation. Instead, they only appear within the quasiclassical belt of states formed by the classes $T$, $X$, and $Q$. Here, the negative outer KD values represent a less apparent form of non-classicality, observable either in weak measurements \cite{Lun11,Hof11,Lun12,Hof12} or in the sensitivity of the system to obstructions or phase shifts in the paths \cite{Han24,Sag25}. For a comprehensive understanding of these non-classical properties of states with negative outer KD values, it is best to consider the $\ket{\theta_k}$ states as representatives of both $Q$ and $X$ class states. Each $\ket{\theta_k}$ state is characterized by a single negative KD value corresponding to the two outer paths connected to the inner path $k$. The five negative KD values read
\begin{eqnarray}
\label{eq:negative}
\varrho(S1,S2|\theta_{3}) &=& - 1/8,
\nonumber \\
\varrho(2,f|\theta_{D1}) &=& - 1/9,
\nonumber \\
 \varrho(1,S2|\theta_{P1}) &=& - 1/10,
\nonumber \\
\varrho(2,S1|\theta_{P2}) &=& - 1/10,
\nonumber \\
\varrho(1,f|\theta_{D2}) &=& - 1/9.
\end{eqnarray} 
The maximal negative value of a KD term $\varrho(i,j)$ is found on the great circle through $\ket{i}$ and $\ket{j}$ between the great circles defined by $P(i)=0$ and $P(j)=0$. Its value is given by
\begin{equation}
 - \varrho(i,j)_{\mathrm{max.}} = \frac{|\braket{i}{j}|(1-|\braket{i}{j}|}{2}.  
\end{equation}
All of the negative KD values given in Eq. (\ref{eq:negative}) are close to their maximum, with $\ket{\theta_3}$ achieving the maximal possible negative KD value of $-1/8$. 

Both the maximal negative value and the maximal positive value of $\varrho(i,j)$ are found on the great circle through $\ket{i}$ and $\ket{j}$. The corresponding states are orthogonal to each other and are found halfway between $\ket{i}$ and $\ket{j}$. As mentioned above, the states with maximal positive $\varrho(i,j)$ are found on the boundary between $B(i,j)$ and $T(i,j)$, which necessarily includes a state orthogonal to $\ket{\theta_k}$. 

The negative KD values of the states $\ket{\theta_k}$ are all outer KD values, that is, both $\ket{i}$ and $\ket{j}$ are outer paths. The maximal negative values of the five outer KD values are equally distributed close to a great circle around the center of the class $N$ pentagon. The states that achieve the maximal positive values for these outer KD values are significantly closer to the center of the clas $N$ pentagon and are hence much closer to each other than the states $\ket{\theta_k}$. 

For the five inner KD values $\varrho(i,l)$, where $\ket{l}$ is an inner path, the situation is opposite. Extremal values are found along a great circle orthogonal to $\ket{\theta_l}$, since $\ket{\theta_l}$ is orthogonal to both $\ket{i}$ and $\ket{l}$. The maximal positive KD value is obtained by a state in sub-class $Q(i,l)$, halfway between $\ket{i}$ and $\ket{l}$. The maximal negative KD value is obtained for a state in class $N$ that is orthogonal to $\ket{\theta_l}$. The maximal negative values are so close together that they all fall into the same class. Oppositely, the maximal positive KD values are spread out near the same great circle where the states $\ket{\theta_k}$ are found. 

The extremal KD values suggest that the sub-classes $Q(i,l)$ and the states $\ket{\theta_k}$ describe a wide variety of distinguishable states characterized by maximal positive inner KD values and maximal negative outer KD values. Each sub-class $Q(i,l)$ maximizes the positive KD value $\varrho(i,l)$ while being characterized by two negative outer KD values. These outer KD values corrspond to the characteristic negative KD values of the states $\ket{\theta_k}$ at the edge of the sub-class $Q(i,l)$. In each of the five contexts, the highest probability is assigned to the paths that lie on a classical trajectory through $\ket{i}$ and $\ket{l}$, consisting of the outer path $\ket{i}$ and three inner paths with $\ket{l}$ at the center.  

The states of sub-class $Q(i,l)$ are approximately orthogonal to the sub-classes that maximize the positive values of the KD values that are negative in $Q(i,l)$. For example, $Q(1,P2)$ is approximately orthogonal to $B(S1,S2)$, $T(S1,S2)$, $B(2,f)$, and $T(2,f)$. These orthogonality relations have a quasiclassical origin, since the trajectory through $\ket{1}$ and $\ket{P2}$ does not share any paths with the trajectory through $\ket{S1}$ and $\ket{S2}$ or the trajectory through $\ket{2}$ and $\ket{f}$. Likewise, the trajectory through $\ket{S1}$ and $\ket{S2}$ does not share any paths with the trajectory through $\ket{2}$ and $\ket{f}$, as evidenced by the negative KD value of $\varrho(2,f)$ in $T(S1,S2)$, and the negative KD value of $\varrho(S1,S2)$ in $T(2,f)$. 

As shown in table \ref{tab1}, each sub-class $T(i,j)$ has two negative inner KD values, suggesting approximate orthogonality to two sub-classes $Q(i,l)$. For $T(S1,S2)$, these approximately orthogonal sub-classes are $Q(1,P2)$ and $Q(2,P1)$. Likewise, $T(2,f)$ is approximately orthogonal to $Q(1,P2)$ and $Q(S1,D2)$. In general, the relations between negative and positive KD values in the classes $Q$ and $T$ can be derived from the classical orthogonality relations between the trajectories represented by the sub-classes, where each sub-class $Q(i,l)$ is orthogonal to two sub-classes $T(i,j)$, and each sub-class $T(i,j)$ is orthogonal to two sub-classes $Q(i,l)$ and two sub-classes $T(i,j)$. The classes $T$, $X$ and $Q$ thus form a quasiclassical belt around the Hilbert space sphere. The class $N$ violates the non-contextual inequality given in Eq. (\ref{eq:ineq}) by being orthogonal to all $Q(i,l)$ at the same time. This is possible because the classes $N$, $V$ and $B$ are much closer together in Hilbert space than the classical outer trajectories represented by class $T$ would allow. The increase in the number of negative inner KD values from class $B$ to class $N$ describes this gradual loss of contrast between the contexts. 

\section{Joint measurements of all five contexts}
\label{Sec:joint}

The $T$ and $Q$ classes of states represent states that minimize the uncertainties of measurement outcomes in all five contexts by maximizing the probabilities of one specific path for each context. The three states at the corners of the sub-classes of $T$ define the states whose probabilities are maximized by the states in that sub-class. States close to the center of the sub-class achieve high fidelities for each of the three states. In the case of the $Q$ sub-classes, the four states needed to cover all five contexts are given by the corners of the quadrilateral formed by $Q(i,l)$ and the neighboring sub-classes $X(i,k)$, where the outer path $\ket{k}$ belongs to two of the three contexts, and the inner paths $\ket{l}$ and$\ket{k}$ belong to the remaining three contexts.  

Negative KD values can be used as indicators of orthogonality relations between the sub-classes of $T$ and $Q$. The sub-classes of $Q$ are characterized by two negative outer KD values, and the maximal positive KD values are found at the boundaries between the corresponding sub-classes of $B$ and $T$. The sub-classes of $T$ also have two negative outer KD values, indicating that they can be orthogonal to each other. For each sub-class of $Q$, it is possible to identify two sub-classes of $T$ from which complete orthogonal basis sets can be constructed. These orthogonal basis sets then describe a simultaneous measurement of path states in all five contexts, where each of the basis states can be interpreted as a minimal uncertainty measurement of the path states at the corners of the sub-group. 

For each sub-class $Q(i,l)$, a suitable state can be found by finding the state orthogonal to both $\ket{N_i}$ and $\ket{\theta_l}$. The non-zero component of $\ket{i}$ in $\ket{N_i}$ responsible for the Hardy-like paradox tilts the result from the state $\ket{i}$ towards the state $\ket{l}$. For example, the state for $Q(S2,D1)$ is given by 
\begin{equation}
\ket{Q(S2,D1)}=\frac{1}{\sqrt{14}}\left(2 \ket{1} - \ket{2} + 3 \ket{3}\right).
\end{equation} 
The negative KD values of this state are $\varrho(2,S1)=-1/14$ and $\varrho(1,f)=-2/21$, indicating that the other two states of the orthogonal basis should be from $T(2,S1)$ and $T(1,f)$. For $T(2,S1)$, we can select the state on the boundary to $B(2,S1)$,
\begin{equation}
\ket{T(2,S1)} = \frac{1}{\sqrt{10}}\left(3 \ket{2} + \ket{3}\right).
\end{equation}
Since this state is orthogonal to $\ket{1}$, it has a KD value of $\varrho(1,f)=0$. It has two negative inner KD values, one of which is $\varrho(S2,D1)=-1/20$, a result of the orthogonality between this state and $\ket{Q(S2,D1)}$. The third state in the orthogonal basis is given by 
\begin{equation}
\ket{T(1,f)} = \frac{1}{\sqrt{35}}\left(5 \ket{1} + \ket{2} - 3\ket{3}\right).
\end{equation}
The orthogonality relations with the other two basis states result in negative KD values of $\varrho(2,S1)=-1/35$ and $\varrho(S2,D1)=-2/35$. 

When implemented as a measurement basis, these three states can identify path states from every context with a fidelity greater than $1/2$. For example, input states of $\{\ket{S1},\ket{f},\ket{P1}\}$ can be discriminated with fidelities of 
\begin{eqnarray}
P(T(2,S1)|S1) &=& 4/5,
\nonumber \\
P(T(1,f)|f) &=& 27/35,
\nonumber \\
P(Q(S2,D1)|P1) &=& 16/21.
\end{eqnarray}
Similar results can be obtained for all other contexts. The three orthogonal states are equally distinguishable in all five contexts of the interferometer. Their orthogonality has a quasiclassical character in the sense that they represent trajectories through the interferometer that do not share any path with each other. States from other sub-classes can be expressed by superpositions of these three states, and the overlap roughly corresponds to the number of contexts in which the trajectories described by the states run along parallel paths. Since the sub-class $T(2,f)$ shares exactly one of its three paths with each of the three basis states, its states can be represented by roughly equal superpositions of the basis states, e.g. 
\begin{eqnarray}
   \ket{T(2,f)} &=& \frac{1}{\sqrt{21}}\left(\ket{1} + 4 \ket{2} – 2 \ket{3}\right)\nonumber \\
   &=& 
\frac{-1}{7 \sqrt{21}} \left(4 \sqrt{14} \ket{Q(S2,D1)} - 7 \sqrt{10} \ket{T(2,S1)} – 3 \sqrt{35} \ket{T(1,f)}\right).
\end{eqnarray}
This state is characterized by probabilities of $P(2|T(2,f))=0.76$, $P(f|T(2,f))=0.78$, and $P(D1|T(2,f))=0.86$. The path states $\ket{2}$, $\ket{f}$ and $\ket{D1}$ all share a considerable overlap with $\ket{T(2,f)}$, yet they are associated with a different outcome of the joint measurement. For comparison,
\begin{eqnarray}
    P(Q(S2,D1)|D1) &=& 0.57 \hspace{0.67cm} P(Q(S2,D1)|T(2,f))= 0.22,
\nonumber \\
  P(T(2,S1)|2) &=& 0.90 \hspace{1cm} P(T(2,S1)|T(2,f))= 0.48,
\nonumber \\
  P(T(1,f)|f) &=& 0.77 \hspace{1.2cm} P(T(1,f)|T(2,f))= 0.31.  
\end{eqnarray}
The measurement fidelity of correctly identifying each of the paths is more than twice as high as the probability of detecting the same outcome in a measurement of the intput state $\ket{T(2,f)}$. The joint measurement is maximally sensitive to the differences between the three path states closest to $\ket{T(2,f)}$, making it ideal for the detection of biases that rotate the state $\ket{T(2,f)}$ on the Hilbert space sphere. 

\section{Conclusions}
\label{Sec:conclude}

The classification system introduced here and illustrated in Fig. \ref{fig2} makes use of both orthogonality and proximity of states in Hilbert space. The boundaries of each sub-class are defined by orthogonality relations with one of the ten path states. The Kirkwood-Dirac quasiprobability is ideal for the characterization of the sub-classes, since crossing a zero probability boundary changes the signs of all KD values associated with the corresponding state. The states violating a non-contextual inequality are characterized by negative values of the five inner KD values, where maximal violations are associated with the $N$ class of states. This class is the only fully symmetric class and has only one sub-class. Moving away from the states with contextual path statistics, classes are subdivided by their non-symmetric relation to the five contexts. A violation of non-contextual inequalities is possible for classes $N$, $V$ and $B$, all of which have only positive outer KD values. States in these classes have $5$, $4$ or $3$ negative inner KD values, with five states $\ket{N_i}$ at the intersections between $N$, $V$ and $B$. These five states are defined by probabilities of zero for two of the inner path, a definition that results in a Hardy-like paradox observed as a non-zero probability in the state $\ket{i}$ that is used to identify the state $\ket{N}_i$. A detailed characterization of these states in terms of the relation between their KD values and the probabilities of finding the photon in different paths gives insights into the quantum statitics of the three classes of states responsible for inequality violations, where the sub-classes of $V$ and $B$ represent an additional degree of freedom gained when moving away from the maximal violation of inequalities observed at the center of class $N$. 

An even greater variety of states is possible when no inequality is violated. This is the case for all states in the quasiclassical belt around the Hilbert space sphere formed by the classes $T$, $X$, and $Q$. The specific statistics of these states vary widely, depending on the respective sub-class. This is reflected by the two negative outer KD values that characterize the sub-classes. These negative KD values now indicate which states in each context have low probabilities, potentially leaving a single preferred path through the interferometer. Maximal negative KD values are found close to the $\ket{\theta_k}$ states located at the intersections of the $X$ and $Q$ classes. As the analysis above has shown, these states have four positive KD values that correspond to an assignment of probabilities to four quasiclassical paths through the interferometer. The maximal values of these four positive KD values are achieved at the center of the $Q$ classes and at the edges of the $T$ classes. Overall, the sub-classes of $Q$ define states that maximize probabilities along a specific trajectory through the interferometer running through one outer state and three inner states, while the sub-classes of $T$ define states that maximize probabilities for a trajectory of two outer paths and one inner path. States in the quasiclassical belt thus represent a joint control of all five contexts with minimal uncertainty. As shown above, a selection of three orthogonal states can be used to realize a joint measurement of all contexts, where quasiclassical states representing trajectories with overlapping paths have a corresponding quantum mechanical overlap. The comprehensive classification of states introduced here thus highlights the onset of quantum-classical correspondence in a three dimensional Hilbert space.

The classification introduced above provides a map to the full range of statistical possibilities of quantum states in a contextual three-path interferometer. The wide variety of sub-classes shows that the increase of Hilbert space dimensions from two to three greatly expands the range of possibilities. Equipped with the map provided in Fig. {\ref{fig2}, we might learn to successfully navigate larger Hilbert spaces in search of new possibilities.

%%%%%%%%%%%%%%%%%%%%%%%%%%%%%%%%%%%%%%%%%%%%%%%%%%%%%%%55

\section*{Acknowledgements}
This work was supported by ERATO, Japan Science and Technology Agency (JPMJER2402).


\vspace{0.5cm}

\begin{thebibliography}{xyz00}

%%%---single photon interference


\bibitem{Wot79}
W. K. Wootters and W. H. Zurek, Complementarity in the
double-slit experiment: Quantum nonseparability and a quan-
titative statement of Bohr’s principle, Phys. Rev. D {\bf 19}, 473 (1979).

\bibitem{Scu91}
M. O. Scully, B.-G. Englert and H. Walther, Quantum optical tests of
complementarity, Nature(London) {\bf 351}, 111 (1991).

\bibitem{Eng96}
B.-G. Englert, Fringe Visibility and Which-Way Information: An Inequality,
Phys. Rev. Lett. {\bf 77}, 2154 (1996).

\bibitem{Kwi97}
P. G. Kwiat, Hyper-entangled states, J. Mod. Opt. {\bf 44}, 2173 (1997).

\bibitem{Dur98}
S. D\"urr, T. Nonn and G. Rempe, Origin of quantum-mechanical complementarity
probed by a ’which-way experiment in an atom interferometer, Nature(London)
{\bf 395}, 33 (1998).

\bibitem{Wal02}
S. P. Walborn, M. O. Terra Cunha, S. Padua, and C. H. Monken, Double-slit quantum eraser, Phys. Rev. A {\bf 65}, 033818 (2002).

%%--qdits and triple slits

\bibitem{Nev05}
L. Neves, G. Lima, J. G. Aguirre Gómez, C. H. Monken, C. Saavedra, and S. Padua,Generation of Entangled States of Qudits using Twin Photons, Phys. Rev. Lett. {\bf 94}, 100501 (2005).

\bibitem{Sid15}
M. A. Siddiqui and T. Qureshi, Three-Slit Interference: A Duality Relation, Prog. Theor. Exp. Phys. {\bf 2015}, 083A02 (2015).

\bibitem{Sha17}
N.A. Shah and T. Qureshi, Quantum eraser for three-slit interference, Pramana-J. Phys. {\bf 89}, 80 (2017).

\bibitem{Vai13}
L. Vaidman, Past of a quantum particle, Phys. Rev. A {\bf 87},
052104 (2013).

\bibitem{Dan13}
A. Danan, D. Farfurnik, S. Bar-Ad, and L. Vaidman, Asking photons where they have been, Phys. Rev. Lett. {\bf 111}, 240402 (2013).

\bibitem{Vai14}
Vaidman, L., Tracing the past of a quantum particle. Phys. Rev. A {\bf 89}, 024102 (2014).


\bibitem{Gri16}
R. B. Griffiths, Particle path through a nested Mach-Zehnder interferometer. Phys. Rev. A {\bf 94}, 032115 (2016).

\bibitem{Eng17}
B. G. Englert, K. Horia, J. Dai, Y.L. Len, and H.K. Ng, Past of a quantum particle revisited. Phys. Rev. A {\bf 96}, 022126 (2017).



\bibitem{Geb18}
H. Geppert-Kleinrath, T. Denkmayr, S. Sponar, H. Lemmel,
T. Jenke, and Y. Hasegawa, Multifold paths of neutrons in the
three-beam interferometer detected by a tiny energy kick, Phys.
Rev. A {\bf 97}, 052111 (2018).

\bibitem{Yua19}
Q. Yuan and X. Feng, Three-path interference of a photon and reexamination of the nested Mach-Zehnder interferometer, Phys. Rev. A {\bf 99}, 053805 (2019).

\bibitem{Spo19}
Sponar, S., Geppert, H., Denkmayr, T., Lemmel, H., Hasegawa, Y., Asking neutrons where they have been, J. Phys.: Conf. Ser. {\bf 1316}, 012002 (2019).

\bibitem{Han23}
Hance, J.R., Rarity, J., Ladyman, J., Weak values and the past of a quantum particle, Phys. Rev. Res. {\bf 5}, 023048 (2023).

\bibitem{Wag24a}
R. Wagner, A. Camillini, and E. F. Galvao, Coherence and contextuality in a Mach-Zehnder interferometer,
Quantum {\bf 8}, 1240 (2024).

%%%---contextuality and scenarios
\bibitem{Spe60}
E. P. Specker, Die Logik Nicht Gleichzeitig Entscheidbarer Aussagen, Dialectica {\bf 14}, 239 (1960) (English translation arXiv:1103.4537).

\bibitem{Koc67}
S. Kochen and E. Specker, The problem of hidden variables in quantum mechanics, Indiana Univ. Math. J. {\bf 17}, 59 (1967).

\bibitem{Can14}
C. Canas, S. Etcheverry, E. S. Gomez, C. Saavedra, G. B. Xavier, G. Lima, and A. Cabello, Experimental implementation of an eight-dimensional Kochen-Specker set and observation of its connection with the Greenberger-Horne-Zeilinger theorem, Phys. Rev. A {\bf 90}, 012119 (2014).

\bibitem{Li17}
T. Li, Q. Zeng, X. Song, and X. Zhang, Experimental Contextuality in Classical Light, Sci. Rep. {\bf 7}, 44467 (2017).

\bibitem{Qu21}
D. Qu, K. Wang, L. Xiao, X. Zhan, and P. Xue, State-independent test of quantum contextuality with either single photons or coherent light, npj Quantum Information {\bf 7}, 154 (2021).

%%%---contextuality in three-dimensional Hilbert spaces

\bibitem{Cli93}
R. Clifton, Getting contextual and nonlocal elements‐of‐reality the easy way, Am. J. Phys. {\bf 61}, 443–447 (1993).


\bibitem{Hua03}
Y.-F. Huang, C.-F. Li, Y.-S. Zhang, J.-W. Pan, and G.-C. Guo, Experimental test of the Kochen-Specker theorem with single photons, Phys. Rev. Lett. {\bf 90}, 250401 (2003).


\bibitem{Lei05}
M. S. Leifer and Robert W. Spekkens, Pre- and post-selection paradoxes and contextuality in quantum mechanics, Phys. Rev. Lett. {\bf 95}, 200405 (2005).

\bibitem{Kly08}
A. A. Klyachko, M. A. Can, S. Binicioglu, and A. S.
Shumovsky, Simple Test for Hidden Variables in Spin-1 Systems, Phys. Rev. Lett. {\bf 101}, 020403 (2008).


\bibitem{Bar09}
H. Bartosik, J. Klepp, C. Schmitzer, S. Sponar, A. Cabello, H. Rauch, and Y. Hasegawa, Experimental Test of Quantum Contextuality in Neutron Interferometry, Phys. Rev. Lett. {\bf 103}, 040403 (2009).


\bibitem{Cab13}
A. Cabello, P. Badziag, M. T. Cunha, and M. Bourennane,
Simple Hardy-Like Proof of Quantum Contextuality, Phys. Rev.
Lett. {\bf 111}, 180404 (2013).

\bibitem{Pav25}
M. Pavicic, Quantum Contextual Hypergraphs, Operators, Inequalities, and Applications in Higher Dimensions, Entropy {\bf 2025}, 27, 54 (2025).


\bibitem{Ji24}
M. Ji and H. F. Hofmann, Quantitative relations between different measurement contexts, Quantum {\bf 8}, 1255 (2024).



%%---

\bibitem{OpticaQ}
H. F. Hofmann, Sequential propagation of a single photon through five measurement contexts in a three-path interferometer, Optica Quantum {\bf 1}, 63
(2023).

%%---Kirkwood-Dirac

\bibitem{Kir33}
J. G. Kirkwood, Phys. Rev. {\bf 44}, 31 (1933).

\bibitem{Dirac}
P. A. M. Dirac, On the Analogy Between Classical and Quantum Mechanics, Rev. Mod. Phys. {\bf 17}, 195 (1945).

\bibitem{Lun11}
J.S. Lundeen, B. Sutherland, A. Patel, C. Stewart and C.Bamber, Direct measurement of the quantum wavefunction, Nature {\bf 474}, 188 (2011).

\bibitem{Hof11}
H. F. Hofmann, On the role of complex phases in the quantum statistics of weak measurements, New J. Phys. {\bf 13}, 103009 (2011).

\bibitem{Lun12}
J.S. Lundeen and C. Bamber, Procedure for direct measurement of general quantum states using weak measurement, Phys. Rev. Lett. {\bf 108}, 070402 (2012).

\bibitem{Hof12}
H. F. Hofmann, Complex joint probabilities as expressions of reversible transformations in quantum mechanics, New J. Phys. {\bf 14}, 043031 (2012).

\bibitem{The17}
G. S. Thekkadath, R.Y. Saaltink, L. Giner, and J. S. Lundeen, Determining complementary properties with quantum clones, Phys. Rev. Lett. 119 050405 (2017).

\bibitem{Hal18}
N. Yunger Halpern, B. Swingle, and J. Dressel, Quasiprobability behind the out-of-time-ordered correlatior, Phys. Rev. A {\bf 97}, 042105 (2018).

\bibitem{Iin18}
M. Iinuma, M. Nakano, H. F. Hofmann, and Y. Suzuki,
Experimental evaluation of the nonclassical relation between
measurement errors using entangled photon pairs as a probe, Phys. Rev. A {\bf 98}, 062109 (2018).

\bibitem{Bud23}
A. Budiyono and H. K. Dipojono,  Quantifying quantum coherence via Kirkwood-Dirac quasiprobability, Phys. Rev. A {\bf 107}, 022408 (2023).


\bibitem{Los23}
M. Lostaglio, A. Belenchia, A. Levy, S. Hernandez-Gomez, N. Fabbri, and S, Gherardini, Kirkwood-Dirac quasiprobability approach to the statistics of incompatible observables, Quantum {\bf 7}, 1128 (2023).

\bibitem{Ume24}
S. Umekawa, J. Lee. and N. Hatano, Advantages of the Kirkwood–Dirac distribution among general quasi-probabilities on finite-state quantum systems, Prog. Theor. Exp. Phys. {\bf 2024}, 023A02 (2024).

\bibitem{Wag24b}
R. Wagner et al., Quantum circuits for measuring weak values, Kirkwood-Dirac quasiprobability distributions, and state spectra, Quantum Science and Technology {\bf 9}, 015030 (2024).

\bibitem{Han24}
J. R. Hance, T. Matsushita, and H.F. Hofmann, Counterfactuality, back-action, and information gain in multi-path interferometers, Quantum Science and Technology {\bf 9}, 045015 (2024).

\bibitem{Arv24}
D.R.M. Arvidsson-Shukur et al., Properties and applications of the Kirkwood-Dirac distribution, New J. Phys. {\bf 26}, 121201 (2024).

%%%--Hardy

\bibitem{Har93}
L. Hardy, Nonlocality for Two Particles Without Inequalities for Almost all Entangled States, Phys. Rev. Lett. {\bf 71}, 1665 (1993).

%%--

\bibitem{Entropy}
H. F. Hofmann, Statistical signatures of quantum contextuality, Entropy {\bf 2024}, 26, 725 (2024).

%%--

\bibitem{Sag25}
Y. Sagawa, J.R. Hance, H. F. Hofmann, and T. Ono, Quantum Contextuality Requires Counterfactual Gain, e-print arXiv:2505.14119 (2025).


\end{thebibliography}
\end{document}